\documentclass[conference]{IEEEtran}

\makeatletter
\def\endthebibliography{%
  \def\@noitemerr{\@latex@warning{Empty `thebibliography' environment}}%
  \endlist
}
\IEEEoverridecommandlockouts
\usepackage{cite}
\usepackage{amsmath,amssymb,amsfonts}
\usepackage{bbm}
\usepackage{graphicx}
\usepackage{textcomp}
\usepackage{url}
\usepackage{xcolor}
\usepackage{kotex}
\usepackage{algorithmicx,algpseudocode}
\usepackage{subfigure}
\usepackage{booktabs}
\usepackage[linesnumbered,ruled]{algorithm2e}
\newcommand{\BfPara}[1]{{\noindent\bf#1.}\xspace}

\def\BibTeX{{\rm B\kern-.05em{\sc i\kern-.025em b}\kern-.08em
    T\kern-.1667em\lower.7ex\hbox{E}\kern-.125emX}}
\begin{document}

\title{Coordinated Multi-Agent Reinforcement Learning for Unmanned Aerial Vehicle Swarms in Autonomous Mobile Access Applications}

\author{
\IEEEauthorblockN{Chanyoung Park}
\IEEEauthorblockA{\textit{Korea University}
\\Seoul, Korea}
\and
\IEEEauthorblockN{Haemin Lee}
\IEEEauthorblockA{\textit{Korea University}
\\Seoul, Korea}
\and
\IEEEauthorblockN{Won Joon Yun}
\IEEEauthorblockA{\textit{Korea University}
\\Seoul, Korea}
\and
\IEEEauthorblockN{Soyi Jung}
\IEEEauthorblockA{\textit{Ajou University}
\\Suwon, Korea}
\and
\IEEEauthorblockN{Joongheon Kim}
\IEEEauthorblockA{\textit{Korea University}
\\Seoul, Korea}
}

\maketitle

\begin{abstract}
This paper proposes a novel centralized training and distributed execution (CTDE)-based multi-agent deep reinforcement learning (MADRL) method for multiple unmanned aerial vehicles (UAVs) control in autonomous mobile access applications. For the purpose, a single neural network is utilized in centralized training for cooperation among multiple agents while maximizing the total quality of service (QoS) in mobile access applications.
\end{abstract}

\section{Introduction}
In order to provide seamless network services in crowded, wild, or extreme areas, which is one of the potential scenarios in 6G networks, the use of unmanned aerial vehicles (UAVs) is widely considered where the UAVs are autonomously operated with deep learning algorithms~\cite{tvt201905shin}. In this case, if multiple UAVs are jointly utilized, they should be coordinated to maximize the performance of multi-UAV cooperation~\cite{tii22yun}. 

In this paper, a multi-agent deep reinforcement learning (MADRL) algorithm is designed and evaluated for autonomous aerial mobile base-station (BS) network coordination where the UAVs take the roles of mobile BSs. In order to achieve our desired goal, one of the promising approaches is \textit{centralized training and distributed execution (CTDE)} where a single neural network architecture is used for training state-action pairs of multiple UAVs in order to maximize mobile BS performance; and then the trained neural network will be shared by UAVs. Then, the individual UAVs work based on the shared neural network in a distributed manner~\cite{tvt202106jung}.

\begin{figure}[t]
    \centering
    \includegraphics[width=1\columnwidth]{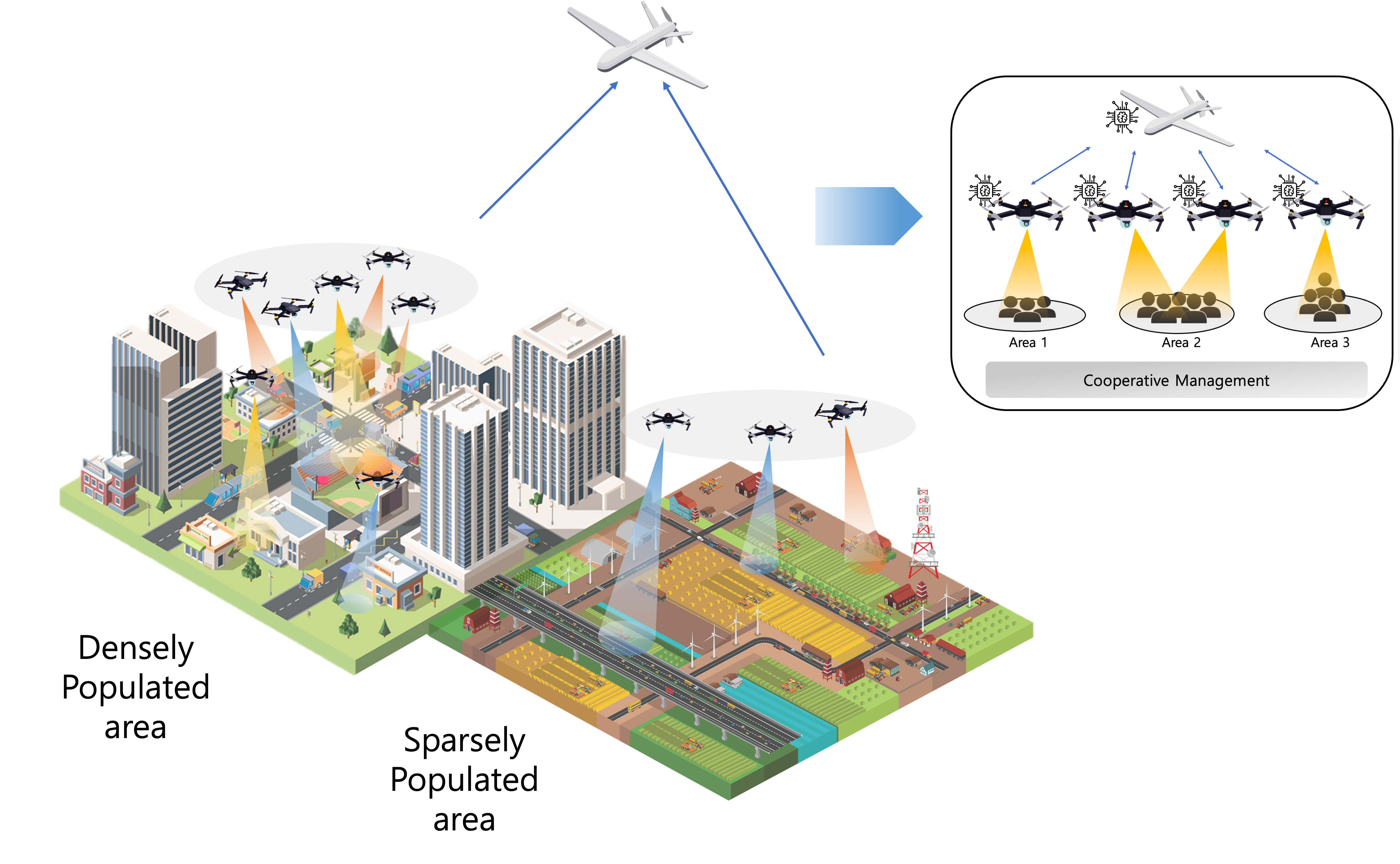}
    \caption{Reference system model for aerial mobile access.}
    \label{fig:system model}
\end{figure}

\section{CTDE-based MADRL for UAV Swarms}

\begin{figure*}[t!]
    \centering
    \includegraphics[width=1.4\columnwidth]{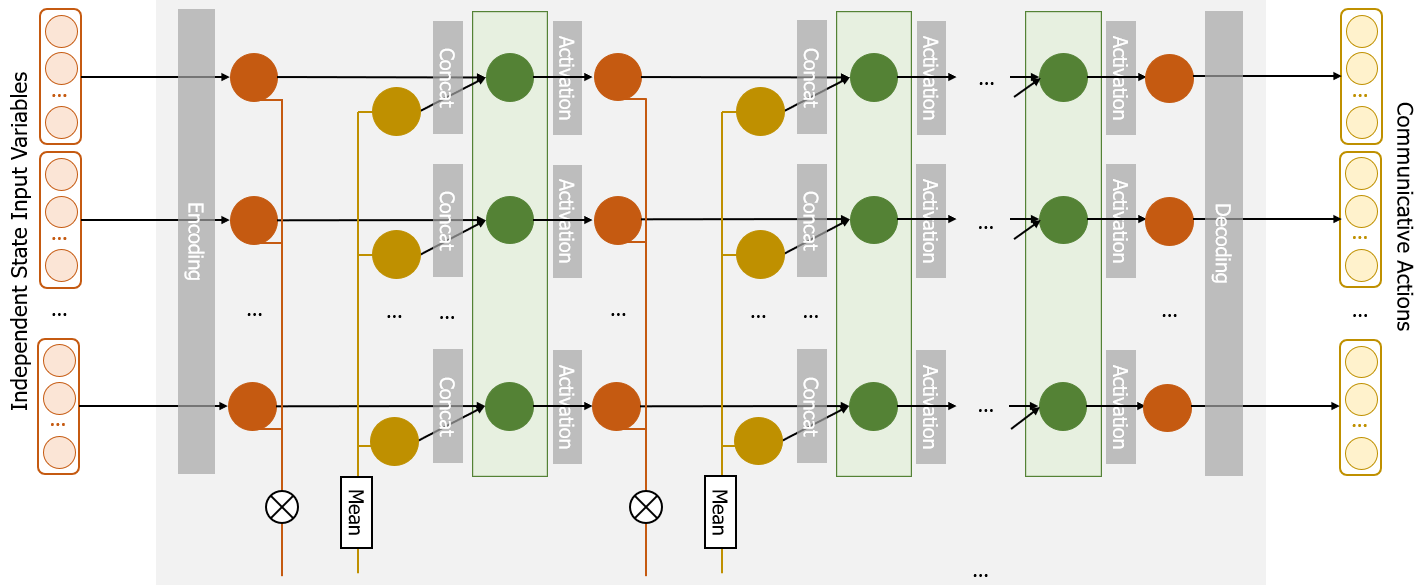}
    \caption{CTDE-based MADRL model for the cooperation and coordination of multiple UAVs.}
    \label{fig:commnet}
    \vspace{-3mm}
\end{figure*}

For CTDE-based MADRL, a single neural network is used to realize cooperation and coordination among multiple UAVs, as shown in Fig.~\ref{fig:commnet}. In the network, multiple states will be inputs (i.e., independent state input variables) and be fed into hidden layers. In the hidden layers, in UAV $i$ among given UAV sets (where $N$ is the set of UAVs, i.e., $N=\{1,\cdots,i,\cdots,|N|\}$), its value (denoted as $v_{i}$) will be added with the average of the other UAVs' state input variables (denoted as $\frac{1}{N-1}\sum_{j\in N, j\neq i}v_{j}$), and the added results will be fed into activation function; then it will be the unit value of next hidden layer~\cite{tii22yun,tvt202106jung}. By iterating this computation multiple times with multiple hidden layers, the states in multiple UAVs can be mixed and jointly considered. In our performance evaluation, we used six hidden layers, showing that the setting is good enough to utilize the desired performance of multi-agent cooperation and coordination. Lastly, at the end of the neural network, a cost function is required, and the function is designed to maximize the quality of services (QoS) in mobile access applications. The reward function in our considering MADRL is designed and implemented for this objective.

\section{Performance Evaluation}
In a real multi-agent environment, each agent can observe different information and only have a limited scope of observation. In this case, CTDE-based MADRL helps multi-agents to cooperatively achieve a common goal in a partially observable Markov decision process (POMDP). However, this is not possible in a real environment due to the limitation of the communication capability of UAVs. Therefore, this paper considers POMDP and validates the performance of our proposed method by comparing it with the performance in a fully observable Markov decision process (FOMDP).

\begin{figure}[t!]
    \centering
    \subfigure[POMDP environment.]{
    \includegraphics[width=0.45\linewidth]{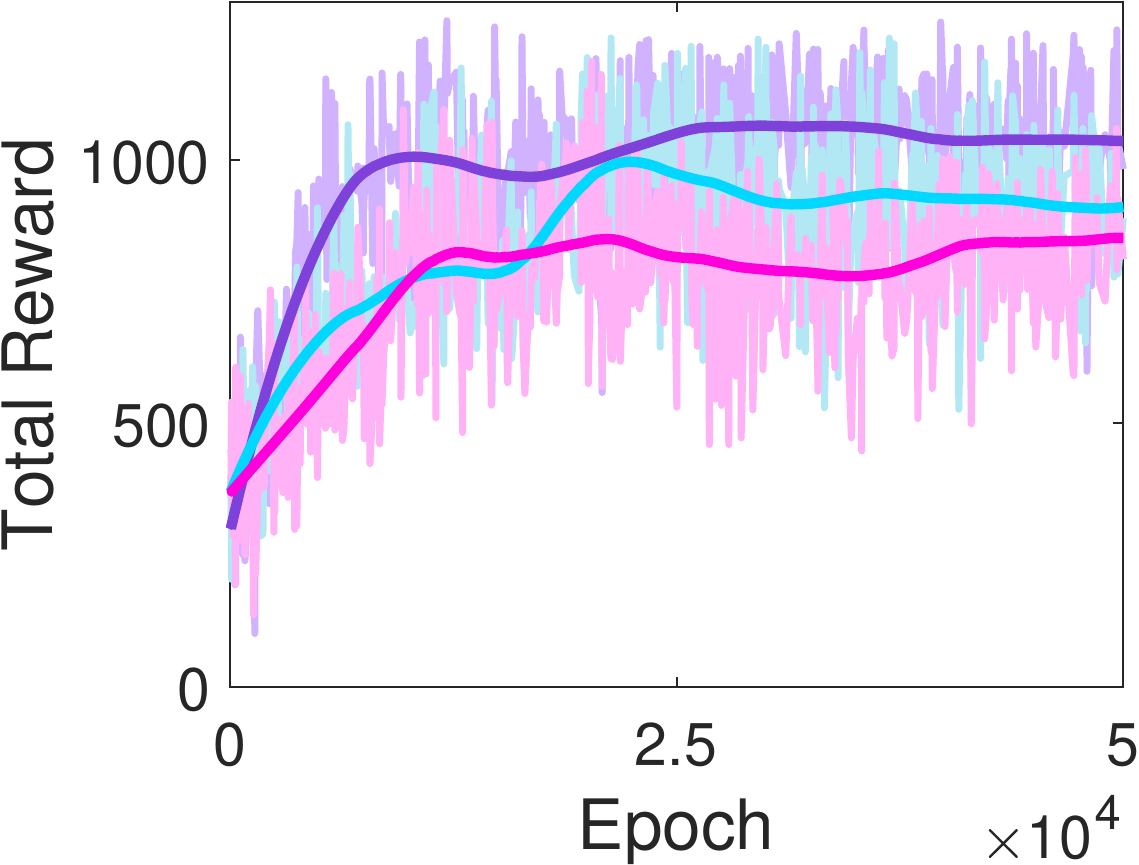}
    \label{fig:POMDP}
    }
    \subfigure[FOMDP environment.]{
    \includegraphics[width=0.45\linewidth]{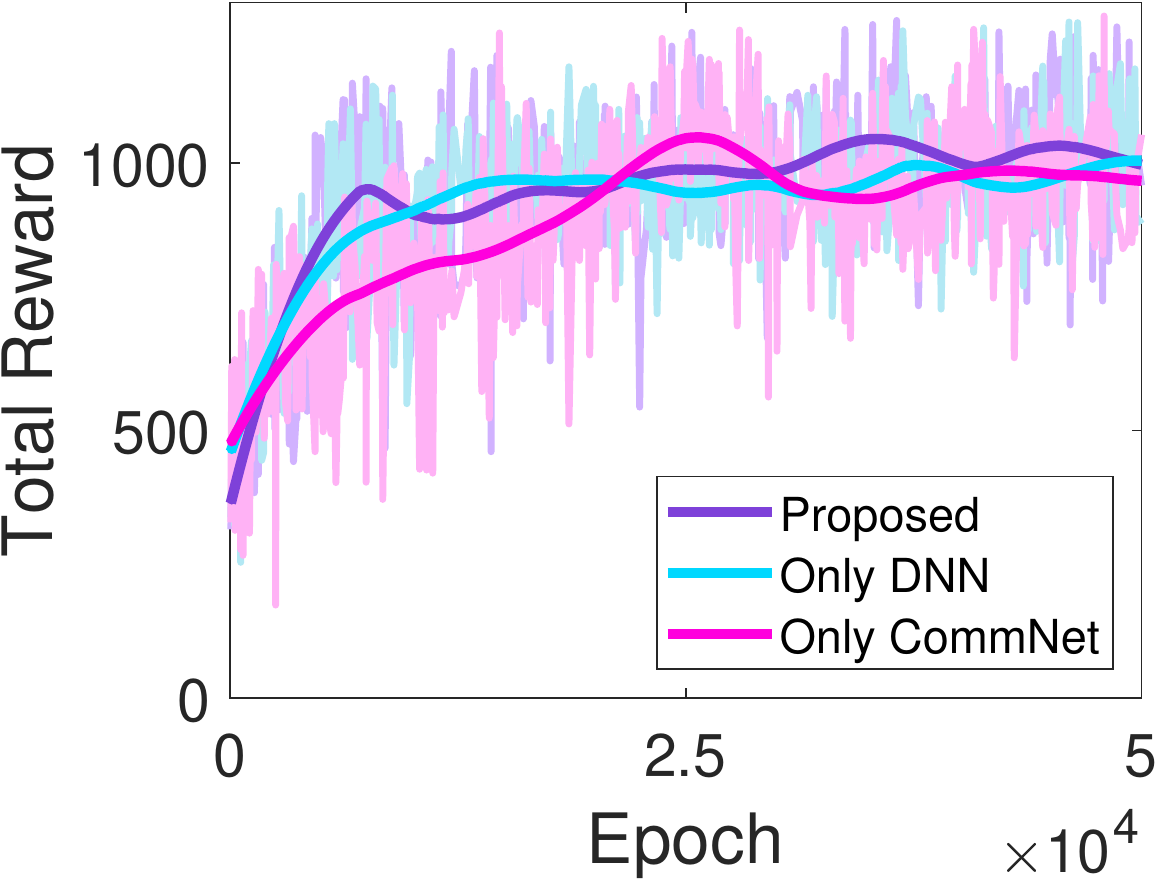}
    \label{fig:FOMDP}
    }\\
    \caption{Total reward in all methods in POMDP.}
    \label{fig:total reward}
\end{figure}

\begin{figure}[ht]
    \centering
    \subfigure[Proposed.]{
    \includegraphics[width=0.45\linewidth]{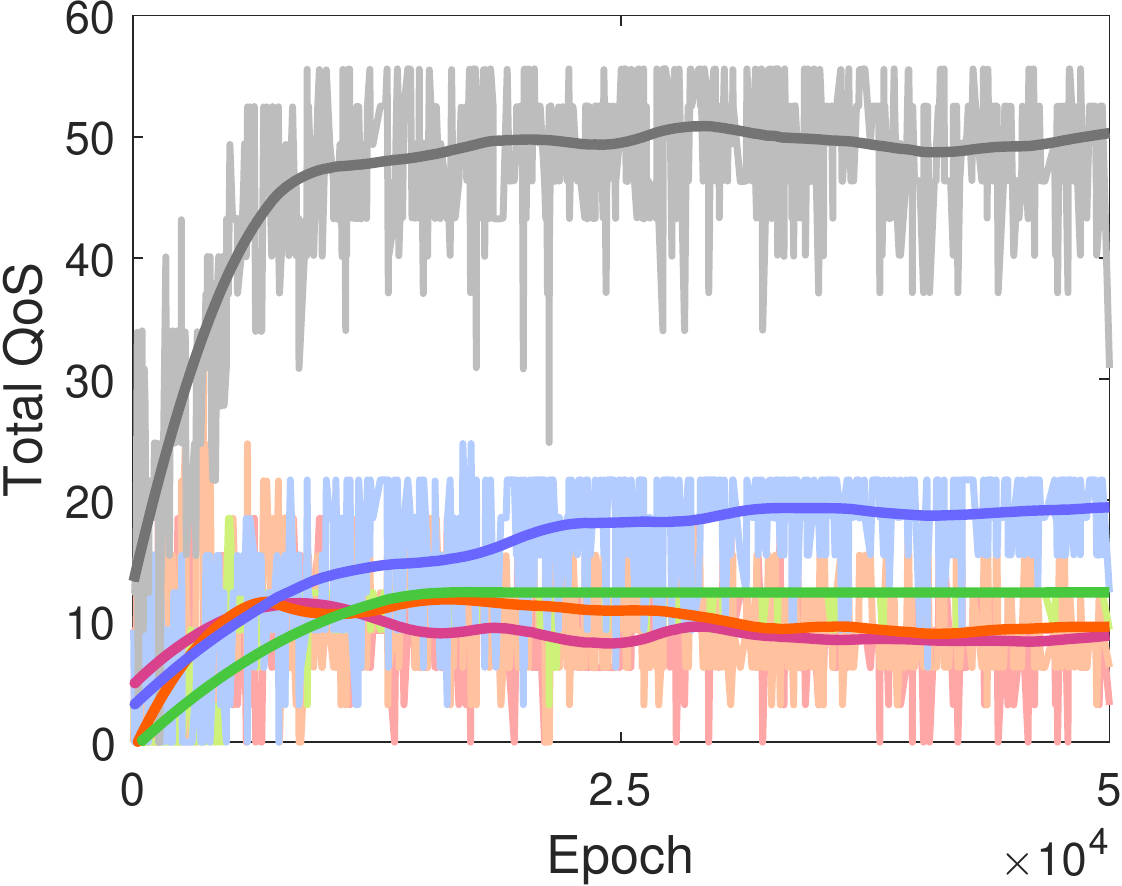}
    \label{fig:quality by Proposed}
    }
    \subfigure[Random.]{
    \includegraphics[width=0.45\linewidth]{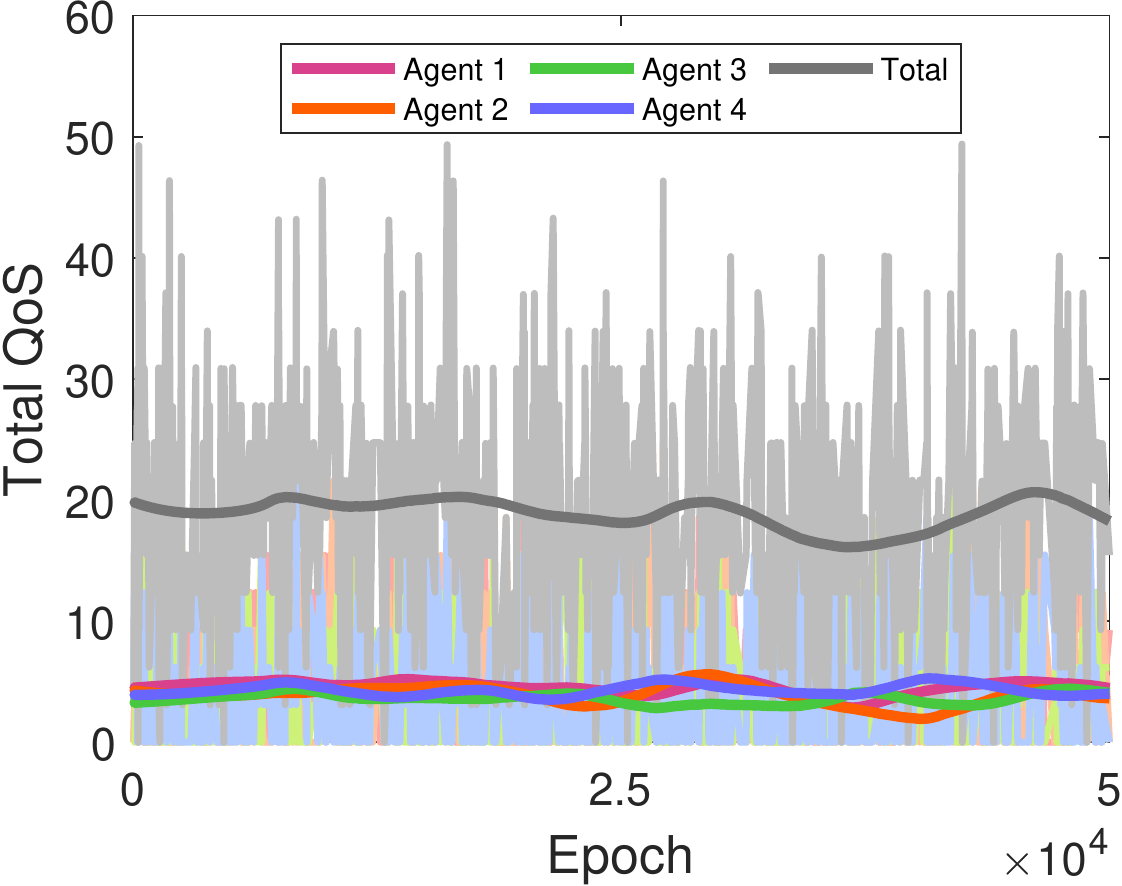}
    \label{fig:quality by Random}
    }\\
    \subfigure[Only DNN.]{
    \includegraphics[width=0.45\linewidth]{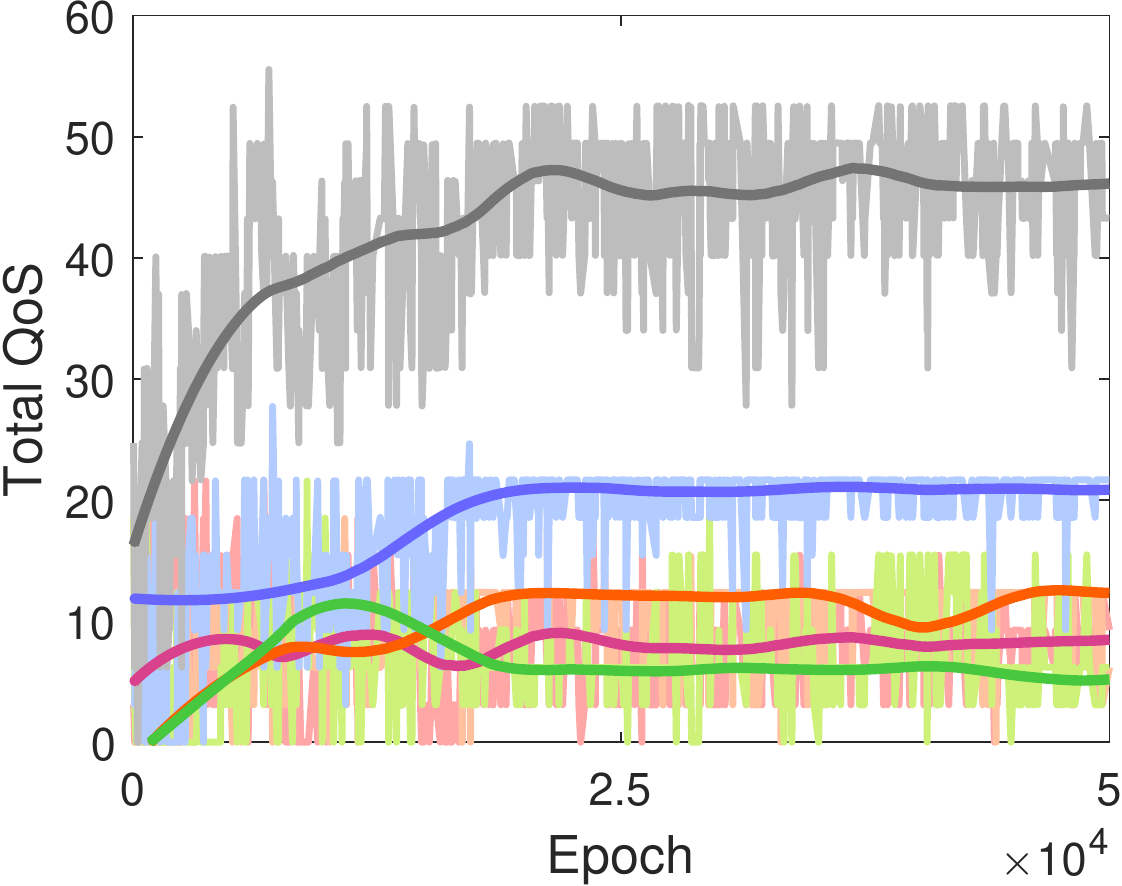}
    \label{fig:quality by Comp1}
    }
    \subfigure[Only CommNet.]{
    \includegraphics[width=0.45\linewidth]{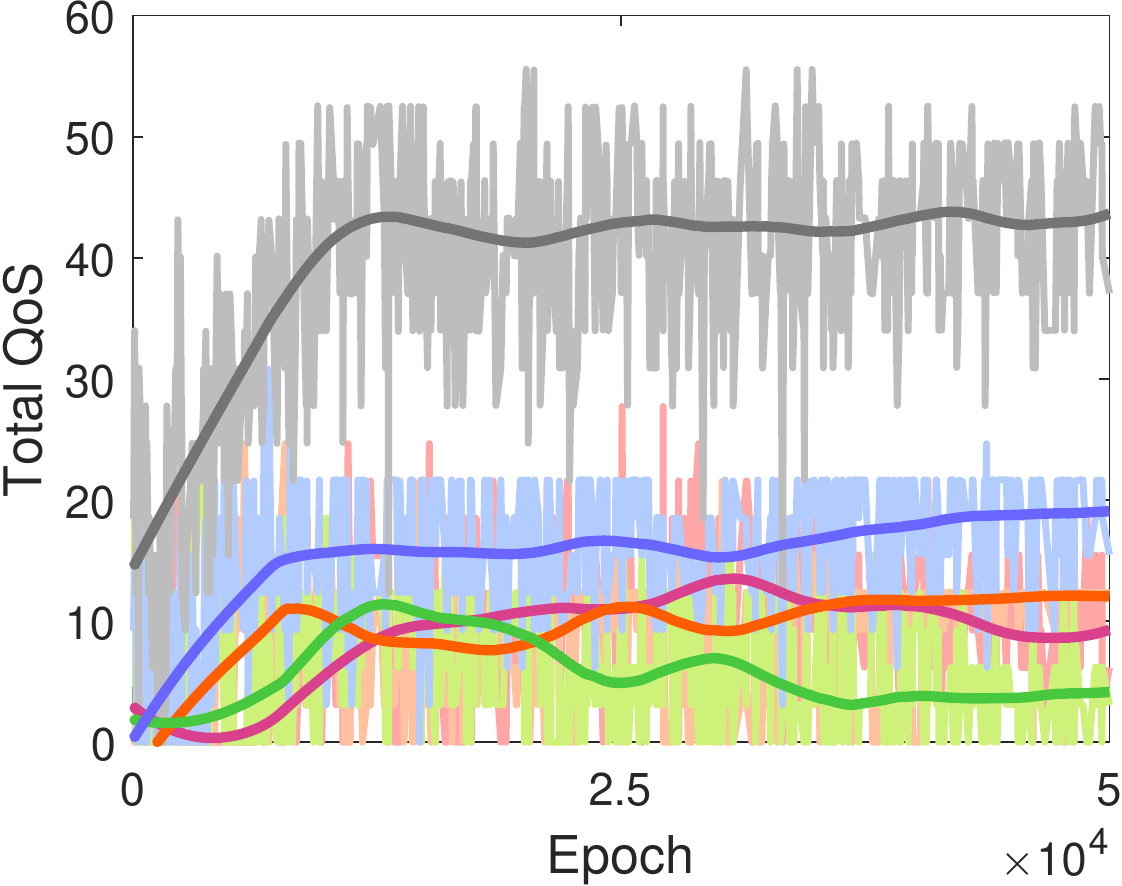}
    \label{fig:quality by Comp2}
    }
    \caption{Each UAV's serving QoS.}
    \label{fig:Quality}
\end{figure}

% Reward Description
Fig.\,\ref{fig:total reward} represents the reward convergence in POMDP and FOMDP, respectively. Our proposed method has one commNet-based leader UAV and three DNN-based non-leader UAVs. In FOMDP, there are no considerations on the limitations of UAV communication capability, thus all methods show similar reward convergence, as presented in Fig.~\ref{fig:FOMDP} (because they do not need to communicate among them to support high-quality services). However, Fig.~\ref{fig:POMDP} shows that our proposed method has the fastest convergence rate in POMDP. In addition, the proposed method shows the best results in terms of reward convergence due to its stability. In a nutshell, our proposed method has the appropriate harmony of inter-UAV communications by the leader UAV and the abundant experiences by non-leader UAVs in a real environment.

% QoS Description
Fig.\,\ref{fig:Quality} shows the QoS received by users in each method. Except for the initial part of training before around 1.6k, users in the proposed method in Fig.\,\ref{fig:quality by Proposed} receive the highest QoS by UAV-BS until the end of the learning. In addition, the proposed method showed the fastest increase rate from 0 to 9k epochs and the most stable convergence rate up to 50 with slight fluctuations. Based on the various experiences of non-leader agents and inter-agent communication of the leader UAV (with a centrally trained neural network), the agent of the proposed method cooperatively provides the highest QoS to Users compared to the Comp1 and Comp2 methods. In the case of agents in the random method in Fig.\,\ref{fig:quality by Random}, the agent delivers QoS evenly to the user. However, the overall QoS of users is very small, so they do not cooperate. Comp1, configured with DNN-based agents in Fig.\,\ref{fig:quality by Comp1}, provides lower QoS than the proposed method because each agent preempts to serve more users. In the Comp2 method composed of CommNet-based agents in Fig.\,\ref{fig:quality by Comp2}, agents try to achieve common goals through communication but cannot provide high QoS with limited information due to the lack of experience.

\section{Conclusions}
This paper proposes a novel MADRL-based multi-UAV swarm control algorithm for autonomous mobile access applications. For the objective, a neural network is used in centralized training for cooperation among multiple UAVs while maximizing the total QoS in mobile services.

\vspace{1mm}
{\small \BfPara{Acknowledgments}
This research was funded by National Research Foundation of Korea (2022R1A2C2004869). Chanyoung Park and Haemin Lee equally contributed to this work (first authors).
Joongheon Kim is a corresponding author (joongheon@korea.ac.kr).}

\bibliographystyle{IEEEtran}  
\bibliography{icdcs23,ref_aimlab}
\end{document}